\documentclass{PoS}

\usepackage{amsmath}
\usepackage{amsfonts}
\usepackage{amssymb}
\usepackage{graphicx}%

\title{Nuclear Lattice Simulations with Chiral Effective Field Theory}

\ShortTitle{Nuclear Lattice Simulations}

\author{\speaker{Dean Lee}\\
        Department of Physics, North Carolina State University, Raleigh, NC 27695, USA\\
        E-mail: \email{dean\_lee@ncsu.edu}}

\abstract{We present recent results on lattice simulations using chiral effective field
theory.  In particular we discuss lattice simulations for dilute neutron
matter at next-to-leading order and three-body forces in light nuclei at
next-to-next-to-leading order.}

\FullConference{8th Conference Quark Confinement and the Hadron Spectrum \\
		 September 1-6 2008\\
		 Mainz, Germany}

\begin{document}

Lattice simulations based on the framework of effective field theory have been
used in studies of nuclear matter \cite{Muller:1999cp} and neutron matter
\cite{Lee:2004qd,Lee:2004si,Abe:2007fe}.
\ The method has also been used to study light nuclei in pionless effective
field theory \cite{Borasoy:2005yc} and chiral effective field theory at
leading order (LO) \cite{Borasoy:2006qn}. \ More recently next-to-leading
order (NLO) calculations have been carried out for the ground state of neutron
matter \cite{Borasoy:2007vi,Borasoy:2007vk}. \ A review of lattice effective
field theory calculations can be found in Ref.~\cite{Lee:2008fa}. \ In this
proceedings article we describe some new results which, at the time of this
writing, are not yet published.

At leading order in chiral effective field theory the nucleon-nucleon
effective potential is%
\begin{equation}
V_{\text{LO}}=V+V_{I^{2}}+V^{\text{OPEP}}.
\end{equation}
$V$, $V_{I^{2}}$ are the two independent contact interactions at leading order
in the Weinberg power counting scheme, and$\ V^{\text{OPEP}}$ is the
instantaneous one-pion exchange potential. \ The interactions in
$V_{\text{LO}}$ can be described in terms of their matrix elements with
two-nucleon incoming and outgoing momentum states. \ For bookkeeping purposes
we label the amplitude as though the two interacting nucleons were
distinguishable, $A$ and $B$. \ In the following $\vec{q}$ denotes the
$t$-channel momentum transfer while $\vec{k}$ is the $u$-channel exchanged
momentum transfer. \ We use $\tau_{I}$ with $I=1,2,3$ to represent Pauli
matrices acting in isospin space and $\sigma_{S}$ with $S=1,2,3$ to represent
Pauli matrices acting in spin space.

For the two leading-order contact interactions the amplitudes are%
\begin{equation}
\mathcal{A}\left(  V\right)  =C,
\end{equation}%
\begin{equation}
\mathcal{A}\left(  V_{I^{2}}\right)  =C_{I^{2}}\sum_{I}\tau_{I}^{A}\tau
_{I}^{B}.
\end{equation}
For the one-pion exchange potential,%
\begin{equation}
\mathcal{A}\left(  V^{\text{OPEP}}\right)  =-\left(  \frac{g_{A}}{2f_{\pi}%
}\right)  ^{2}\frac{\sum_{I}\tau_{I}^{A}\tau_{I}^{B}\sum_{S}q_{S}\sigma
_{S}^{A}\sum_{S^{\prime}}q_{S^{\prime}}\sigma_{S^{\prime}}^{B}}{q^{\,2}%
+m_{\pi}^{2}}.
\end{equation}
For our physical constants we take $m=938.92$ MeV as the nucleon mass,
$m_{\pi}=138.08$ MeV as the pion mass, $f_{\pi}=93$ MeV as the pion decay
constant, and $g_{A}=1.26$ as the nucleon axial charge.

In Ref.~\cite{Borasoy:2006qn} two different lattice actions were considered
which were later denoted LO$_{1}$ and LO$_{2}$ \cite{Borasoy:2007vi}. \ The
interactions in $V_{\text{LO}_{1}}$ include one-pion exchange and two
zero-range contact interactions corresponding with amplitude%
\begin{equation}
\mathcal{A}\left(  V_{\text{LO}_{1}}\right)  =C+C_{I^{2}}\sum_{I}\tau_{I}%
^{A}\tau_{I}^{B}+\mathcal{A}\left(  V^{\text{OPEP}}\right)  .
\end{equation}
The interactions in $V_{\text{LO}_{2}}$ consist of one-pion exchange and two
Gaussian-smeared contact interactions,%
\begin{equation}
\mathcal{A}\left(  V_{\text{LO}_{2}}\right)  =Cf(\vec{q})+C_{I^{2}}f(\vec
{q})\sum_{I}\tau_{I}^{A}\tau_{I}^{B}+\mathcal{A}\left(  V^{\text{OPEP}%
}\right)  ,
\end{equation}
where $f(\vec{q})$ is a lattice approximation to a Gaussian function. \ The
smeared\ interactions in LO$_{2}$ are used to better reproduce $S$-wave phase
shifts for nucleon momenta up to the pion mass. \ The coefficients $C$ and
$C_{I^{2}}$ are tuned to reproduce the physical $S$-wave scattering lengths.
\ In Ref.~\cite{Borasoy:2007vi} nucleon-nucleon phase shifts were calculated for these two
lattice actions using the spherical wall method \cite{Borasoy:2007vy}
at\ spatial lattice spacing $a=(100$ MeV$)^{-1}$ and temporal lattice spacing
$a_{t}=(70$ MeV$)^{-1}$. \ For each case NLO corrections were also computed 
perturbatively and
the unknown operator coefficients determined by fitting to low-energy scattering data.

In Ref.~\cite{Borasoy:2007vk} the ground state energy for dilute neutron matter
was computed using the lattice action LO$_{2}$ and auxiliary-field Monte
Carlo. \ Next-to-leading-order corrections to the energy were also calculated
perturbatively.
\ In this calculation the largest source of systematic error was the large
size of NLO corrections for Fermi momenta larger than $100$ MeV. \ This 
was due to attractive $P$-wave interactions generated by Gaussian smearing
in LO$_{2}$ that needed to be cancelled at next-to-leading order. \ In systems with both
protons and neutrons this $P$-wave correction is\ numerically small when
compared with the strong binding produced by $S$-wave interactions. \ For pure
neutron matter, however, the $S$-wave interactions produce much less binding
due to Fermi repulsion. \ Therefore on a relative scale, the $P$-wave
interactions are not as small an effect in neutron matter.

\begin{figure}
[ptb]
\begin{center}
\includegraphics[
height=2.8037in,
width=3.1522in
]%
{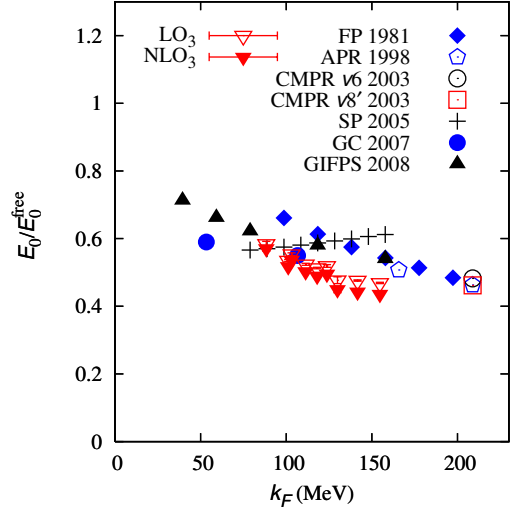}%
\caption{Ground state energy ratio $E_{0}/E_{0}^{\text{free}}$ for LO$_{3}$
and NLO$_{3}$ versus Fermi momentum $k_{F}$. \ For comparison we show results
for FP 1981 \cite{Friedman:1981qw}, APR 1998 \cite{Akmal:1998cf}, CMPR $v6$
and $v8^{\prime}$ 2003 \cite{Carlson:2003wm}, SP 2005 \cite{Schwenk:2005ka},
GC 2007 \cite{Gezerlis:2007fs}, and GIFPS 2008 \cite{Gandolfi:2008id}.}%
\label{kf_literature}%
\end{center}
\end{figure}

These problems have been resolved using a new leading-order action LO$_{3}$
\cite{Epelbaum:2008a}. \ The interactions in $V_{\text{LO}_{3}}$ correspond
with the amplitude,%
\begin{align}
\mathcal{A}\left(  V_{\text{LO}_{3}}\right)   &  =C_{S=0,I=1}f(\vec{q})\left(
\frac{1}{4}-\frac{1}{4}\sum_{S}\sigma_{S}^{A}\sigma_{S}^{B}\right)  \left(
\frac{3}{4}+\frac{1}{4}\sum_{I}\tau_{I}^{A}\tau_{I}^{B}\right) \nonumber\\
&  +C_{S=1,I=0}f(\vec{q})\left(  \frac{3}{4}+\frac{1}{4}\sum_{S}\sigma_{S}%
^{A}\sigma_{S}^{B}\right)  \left(  \frac{1}{4}-\frac{1}{4}\sum_{I}\tau_{I}%
^{A}\tau_{I}^{B}\right)  +\mathcal{A}\left(  V^{\text{OPEP}}\right)  .
\end{align}
The Gaussian-smeared interactions are multiplied by spin and isospin
projection operators. \ Only the $C_{S=0,I=1}$ term contributes in pure
neutron matter. \ Using the LO$_{3}$ action with NLO corrections, we have
computed the ground state energy for dilute neutrons in a periodic box
\cite{Epelbaum:2008a}. \ For\ spatial lattice spacing $a=(100$ MeV$)^{-1}$ and
temporal lattice spacing $a_{t}=(70$ MeV$)^{-1}$ simulations were done with
$8,$ $12,$ $16$ neutrons in periodic boxes with lengths $L=4,5,6,7$. \ In
Fig.~\ref{kf_literature} we show results for the ratio of the interacting
ground state energy to non-interacting ground state energy, 
$E_{0}/E_{\text{0}}^{\text{free}}$, as a function of Fermi momentum $k_{F}$. \ For
comparison we show other results from the literature: \ FP 1981
\cite{Friedman:1981qw}, APR 1998 \cite{Akmal:1998cf}, CMPR $v6$ and
$v8^{\prime}$ \cite{Carlson:2003wm}, SP 2005 \cite{Schwenk:2005ka}, GC 2007
\cite{Gezerlis:2007fs}, and GIFPS 2008 \cite{Gandolfi:2008id}.%

\begin{figure}
[ptb]
\begin{center}
\includegraphics[
height=2.5313in,
width=3.4108in
]%
{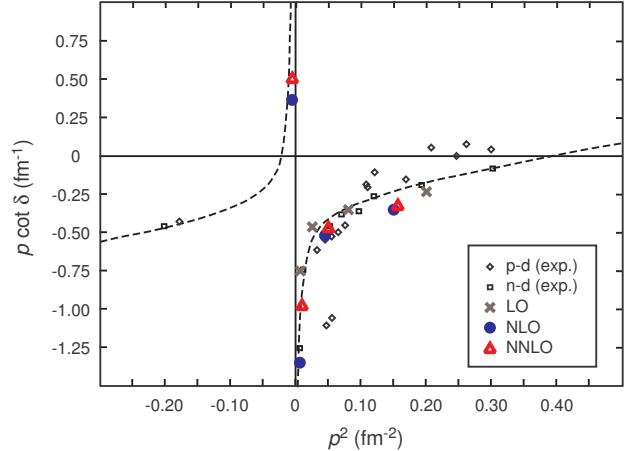}%
\caption{Results for the spin-doublet nucleon-deuteron scattering phase
shift at LO, NLO, and NNLO.}%
\label{nd_doublet}%
\end{center}
\end{figure}

At next-to-next-to-leading order (NNLO) in chiral effective field theory we find
contributions due to three-nucleon forces. \ These interactions consist of a pure
contact interaction, one-pion exchange, and two-pion exchange
\cite{Epelbaum:2002vt}. \ The coupling of one or more pions to a single
nucleon is constrained by chiral symmetry and the corresponding low energy
constants are known \cite{Bernard:1995dp}. \ In the limit of exact isospin
symmetry there are only two unknown coefficients, one for the three-nucleon
contact interaction and one for the two-nucleon-pion vertex involved in the
one-pion exchange interaction. \ At fixed lattice spacing we have determined
these two unknown coefficients by fitting to the triton binding energy and
spin-doublet nucleon-deuteron scattering phase shifts via L\"{u}scher's finite
volume formula \cite{Luscher:1985dn}. \ Results
for the doublet nucleon-deuteron scattering phase shift are shown in
Fig.~\ref{nd_doublet} using the LO$_{2}$ lattice action for lattice spacing
$a=(100$ MeV$)^{-1}$ and temporal lattice spacing $a_{t}=(150$ MeV$)^{-1}$
\cite{Epelbaum:2008a}.%

Having determined the NNLO three-body forces, we have computed the ground
state of the alpha particle without Coulomb interactions on a periodic lattice
using auxiliary-field projection Monte Carlo \cite{Epelbaum:2008a}.  The
NNLO results are within $5\%$ of the actual Coulomb-subtracted alpha binding
energy of about $29$ MeV. \ This is consistent with the expected size of
errors for our chosen lattice spacing and order in effective field theory.

\section*{Acknowledgements}

The work presented here was done in collaboration with Bugra Borasoy, Evgeny
Epelbaum, Hermann Krebs, and Ulf-G. Mei\ss ner. \ Partial financial support
from the Deutsche Forschungsgemeinschaft (SFB/TR 16), Helmholtz Association
(contract number VH-NG-222 and VH-VI-231), and U.S. Department of Energy
(DE-FG02-03ER41260) are acknowledged. \ This research is part of the EU
Integrated Infrastructure Initiative in Hadron Physics under contract number
RII3-CT-2004-506078. \ The computational resources for this project were
provided by the J\"{u}lich Supercomputing Centre at the Forschungszentrum J\"{u}lich.

\end{document}